\documentclass[twocolumn,showpacs,floats,prl,aps,unsortedaddress,superscriptaddress]{revtex4-1}
\usepackage[dvips]{graphicx}
\usepackage{amsfonts}
\usepackage{amsmath}
\usepackage{amssymb}
\usepackage{exscale}
\usepackage{eufrak}
\usepackage{afterpage}
\usepackage{upgreek}
\usepackage{color}	
\usepackage{braket}	
\usepackage{multirow}	

\usepackage{epstopdf}

\newcommand{\srvo}{SrVO$_3$}

\usepackage[normalem]{ulem}    

\begin{document}

\title{Hubbard band or oxygen vacancy states
in the correlated electron metal {\srvo}?}

\author{S.~Backes}
\thanks{S.B. and T.C.R. contributed equally to this work}
\affiliation{Institut f\"ur Theoretische Physik, Goethe-Universit\"at Frankfurt, 
			 Max-von-Laue-Str. 1, 60438 Frankfurt am Main, Germany}
			 
\author{T.~C.~R\"odel}
\thanks{S.B. and T.C.R. contributed equally to this work}
\affiliation{CSNSM, Universit\'e Paris-Sud and CNRS/IN2P3, B\^atiments 104 et 108, 
			91405 Orsay cedex, France}
\affiliation{Synchrotron SOLEIL, L'Orme des Merisiers, Saint-Aubin-BP48, 
			91192 Gif-sur-Yvette, France}

\author{F.~Fortuna}
\affiliation{CSNSM, Universit\'e Paris-Sud and CNRS/IN2P3, B\^atiments 104 et 108, 
			91405 Orsay cedex, France}

\author{E.~Frantzeskakis}
\affiliation{CSNSM, Universit\'e Paris-Sud and CNRS/IN2P3, B\^atiments 104 et 108, 
			91405 Orsay cedex, France}

\author{P.~Le~F\`evre}
\affiliation{Synchrotron SOLEIL, L'Orme des Merisiers, Saint-Aubin-BP48, 
			91192 Gif-sur-Yvette, France}

\author{F.~Bertran}
\affiliation{Synchrotron SOLEIL, L'Orme des Merisiers, Saint-Aubin-BP48, 
			91192 Gif-sur-Yvette, France}

\author{M.~Kobayashi}
\affiliation{Photon Factory, Institute of Materials Structure Science,
			High Energy Accelerator Research Organization (KEK), 
			1-1 Oho, Tsukuba 305-0801, Japan}

\author{R.~Yukawa}
\affiliation{Photon Factory, Institute of Materials Structure Science,
			High Energy Accelerator Research Organization (KEK), 
			1-1 Oho, Tsukuba 305-0801, Japan}

\author{T.~Mitsuhashi}
\affiliation{Photon Factory, Institute of Materials Structure Science,
			High Energy Accelerator Research Organization (KEK), 
			1-1 Oho, Tsukuba 305-0801, Japan}
			
\author{M.~Kitamura}
\affiliation{Photon Factory, Institute of Materials Structure Science,
			High Energy Accelerator Research Organization (KEK), 
			1-1 Oho, Tsukuba 305-0801, Japan}

\author{K.~Horiba}
\affiliation{Photon Factory, Institute of Materials Structure Science,
			High Energy Accelerator Research Organization (KEK), 
			1-1 Oho, Tsukuba 305-0801, Japan}

\author{H.~Kumigashira}
\affiliation{Photon Factory, Institute of Materials Structure Science,
			High Energy Accelerator Research Organization (KEK), 
			1-1 Oho, Tsukuba 305-0801, Japan}

\author{R.~Saint-Martin}
\affiliation{SP2M-ICMMO, CNRS UMR-8182, Universit\'e Paris-Sud, 91405 Orsay Cedex, France}

\author{A.~Fouchet}
\affiliation{GEMaC, CNRS UMR-8635, Universit\'e de Versailles St. Quentin en Yvelines, 
			45 avenue des Etats-Unis, 78035 Versailles Cedex, France}

\author{B.~Berini}
\affiliation{GEMaC, CNRS UMR-8635, Universit\'e de Versailles St. Quentin en Yvelines, 
			45 avenue des Etats-Unis, 78035 Versailles Cedex, France}

\author{Y.~Dumont}
\affiliation{GEMaC, CNRS UMR-8635, Universit\'e de Versailles St. Quentin en Yvelines, 
			45 avenue des Etats-Unis, 78035 Versailles Cedex, France}

\author{A.~J.~Kim}
\affiliation{Institut f\"ur Theoretische Physik, Goethe-Universit\"at Frankfurt, 
			Max-von-Laue-Str. 1, 60438 Frankfurt am Main, Germany}

\author{F.~Lechermann}
\affiliation{Institut f\"ur Theoretische Physik, Universit\"at Hamburg, Jungiusstrasse 9, 
			20355 Hamburg, Germany}

\author{H.~O.~Jeschke}
\affiliation{Institut f\"ur Theoretische Physik, Goethe-Universit\"at Frankfurt, 
			Max-von-Laue-Str. 1, 60438 Frankfurt am Main, Germany}

\author{M.~J.~Rozenberg}
\affiliation{Laboratoire de Physique des Solides, Universit\'e Paris-Sud,
			B\^atiment 510, 91405 Orsay, France}

\author{R.~Valent\'{\i}}
\email{valenti@itp.uni-frankfurt.de}
\affiliation{Institut f\"ur Theoretische Physik, Goethe-Universit\"at Frankfurt, 
			Max-von-Laue-Str. 1, 60438 Frankfurt am Main, Germany}

\author{A.~F.~Santander-Syro}
\email{andres.santander@csnsm.in2p3.fr}
\affiliation{CSNSM, Universit\'e Paris-Sud and CNRS/IN2P3, B\^atiments 104 et 108, 
			91405 Orsay cedex, France}

\date{\today}


\begin{abstract}
We study the effect of oxygen vacancies on the electronic structure 
of the model strongly correlated metal SrVO$_3$. 
By means of angle-resolved photoemission (ARPES) synchrotron experiments,
we investigate the systematic effect of the UV dose on the measured spectra. 
We observe the onset of a spurious dose-dependent prominent peak
at an energy range were the lower Hubbard band has been previously reported in this compound, 
raising questions on its previous interpretation. 
By a careful analysis of the dose dependent effects we succeed
in disentangling the contributions coming from the oxygen vacancy states 
and from the lower Hubbard band.  
We obtain the {\it intrinsic} ARPES spectrum for the zero-vacancy limit, 
where a clear signal of a lower Hubbard band remains. 
We support our study by means of state-of-the-art {\it ab initio} calculations 
that include correlation effects and the presence of oxygen vacancies. 
Our results underscore the relevance of potential spurious states
affecting ARPES experiments in correlated metals, 
which are associated to the ubiquitous oxygen vacancies as extensively
reported in the context of a two-dimensional electron gas (2DEG) 
at the surface of insulating $d^0$ transition metal oxides.
\end{abstract}
%
\maketitle

\begin{figure}[tb]
        \includegraphics[clip, width=0.25\textwidth]{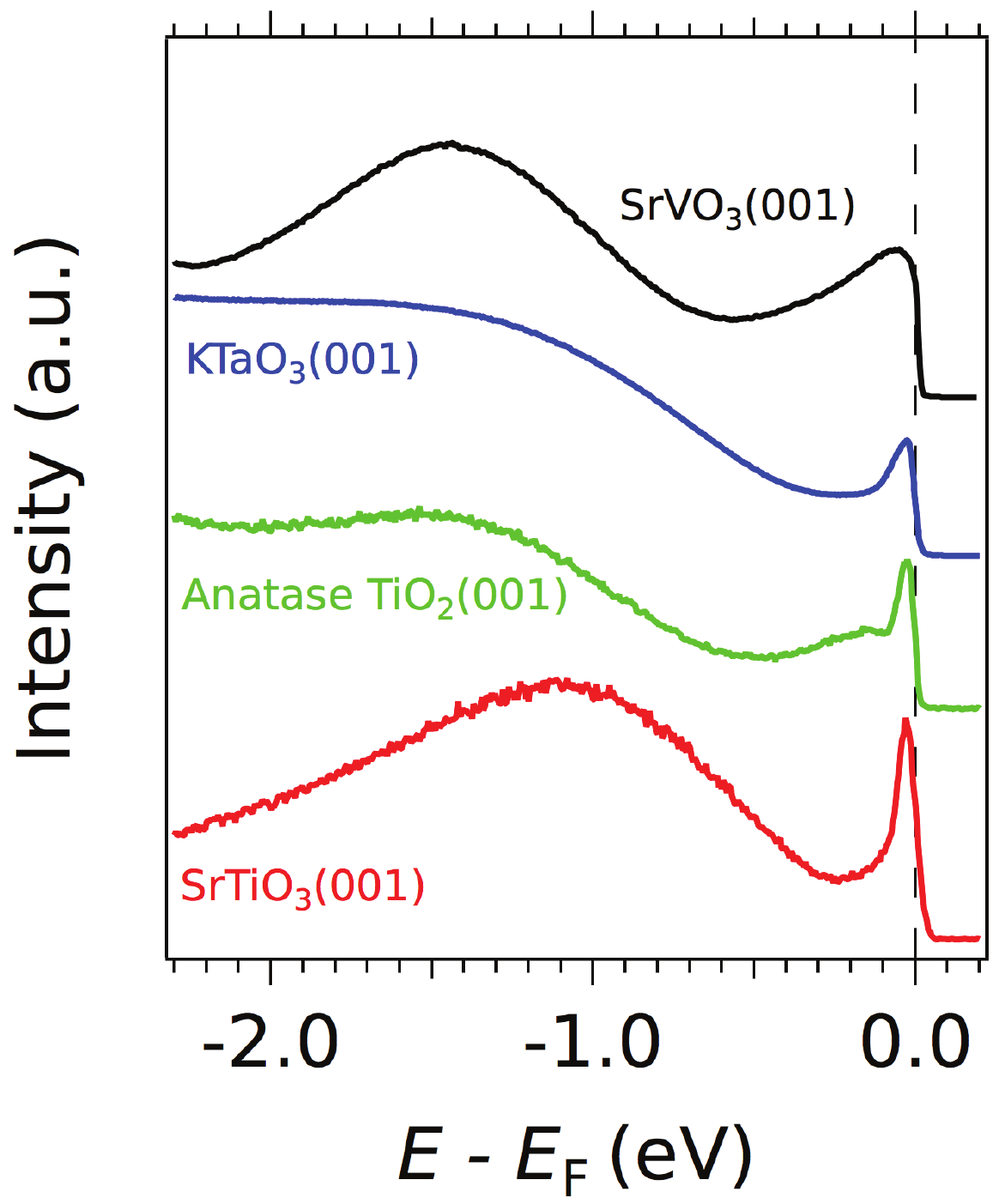}
    \caption{\label{fig:CompSTO-KTO-TiO2-SVO} {(Color online)
        Integrated UV photoemission spectra for various perovskite
        oxides, showing a quasiparticle peak at $E_{\text{F}}$ and an in-gap
        state at energies between 1~eV and 1.5~eV.
        For SrVO$_3$ (upper black curve), a correlated-electron metal, the QP peak
        corresponds to the bulk conduction band, and as will be shown
        further, the in-gap sate is a superposition of the lower
        Hubbard band and localized electronic states associated to
        oxygen vacancies.
        For the other $d^0$ oxides, such as KTaO$_3$ (blue curve), 
        anatase TiO$_2$ (green curve), or SrTiO$_3$ (red curve),
        the QP peak and in-gap state correspond respectively 
        to a confined quasi-2D electron gas at
        the sample surface and to localized states, all formed by
        oxygen vacancies.  The crystal orientation (normal to the
        samples' surface) is indicated in all cases.
        }
      }
\end{figure}

\begin{figure}[tb]
        \includegraphics[width=0.45\textwidth]{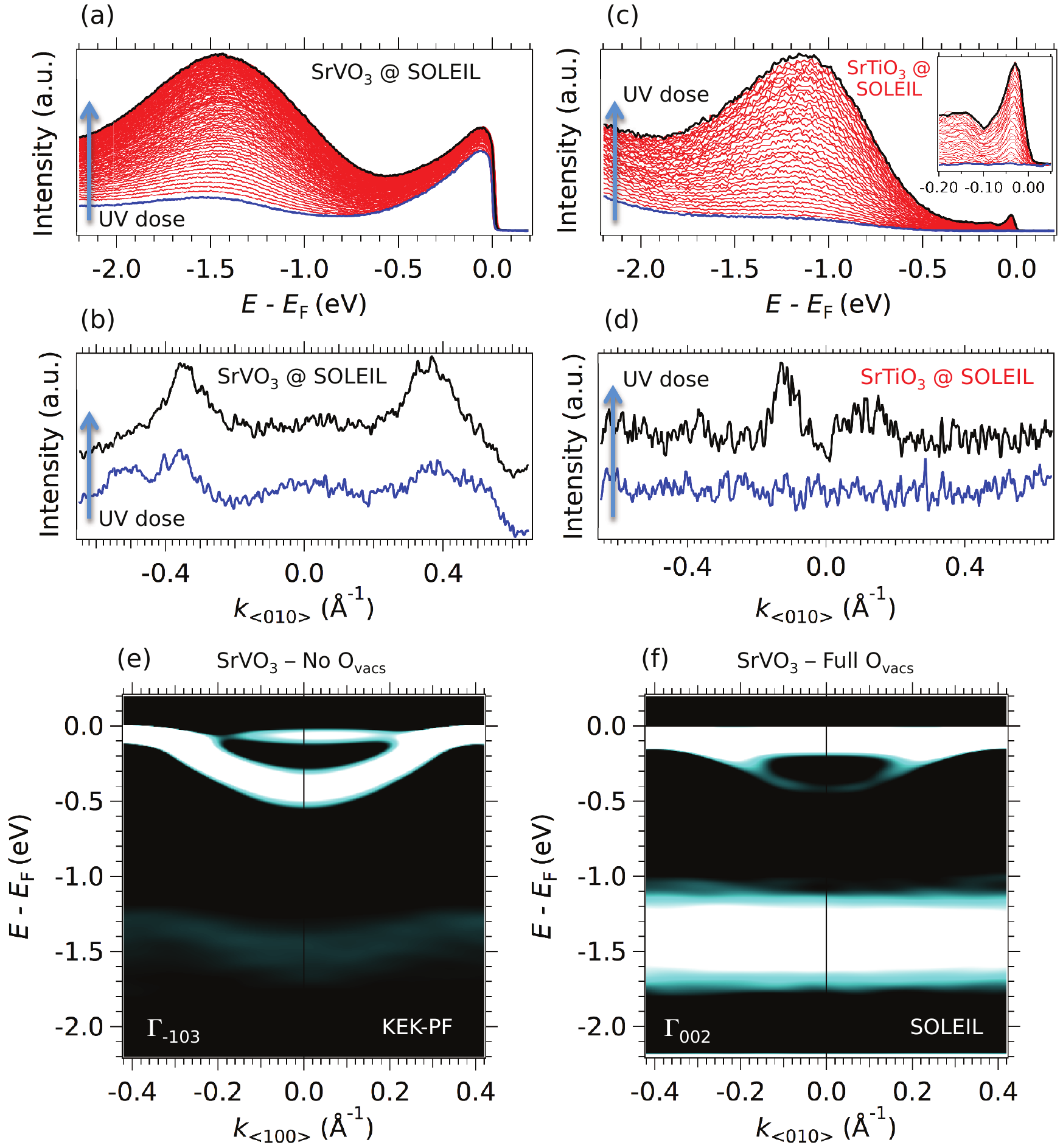}
    \caption{\label{fig:SVO-vs-STO-UVdose} {(Color online)
        (a)~Photoemission spectra of SrVO$_3$ as a function of UV
        dose, measured at Synchrotron SOLEIL.  
        The energy distribution curves (EDCs) were extracted
        from raw ARPES data around the $\Gamma_{002}$ point integrated
        along the $k = <010>$ direction. 
        (b)~Corresponding momentum distribution curves (MDCs)
        integrated over 50~meV below $E_{\text{F}}$.  Peaks in the MDCs
        indicate the Fermi momenta.  
        (c,~d)~Same as (a,~b) for SrTiO$_3$. The filling of a 2DEG 
        upon UV irradiation is evidenced by the formation of QP peaks 
        in the EDCs and MDCs at $E_{\text{F}}$ (inset of (c) and panel (d), respectively).
        (e)~Energy-momentum ARPES intensity map (second derivative,
        negative values) measured at KEK-PF
        with a low UV dose on an SrVO$_3$ sample prepared \emph{in-situ},
        using a well-established protocol to minimize the formation of oxygen vacancies
        (see main text and Supplemental Material).
        The use of second derivatives in the ARPES data allows a better visualization 
        of the dispersion of both the quasi-particle and the Mott-Hubbard bands 
        on the same color plot. The dispersionless feature at $E_{\text{F}}$ 
        is a spurious effect of such
        second-derivative on the Fermi-Dirac cutoff.
        Note that due to the choice of light polarization,
        the heavy bands along $(100)$ are not observed and only
        the contribution of the light $d_{xy}$ band is detected. 
        (f)~Same as (e) after a strong UV irradiation dose, 
        measured at SOLEIL,
        and typical of modern third-generation synchrotrons.
        The larger amount of vacancies enhances the background 
        of inelastic electron scattering, as seen from panel (a), 
        hence increasing the height of the Fermi-Dirac step and the intensity 
        of the spurious feature at $E_{\text{F}}$.
        The raw spectra for panels (e) and (f)
        is shown in the Supplemental Material.
        All data were taken at 20~K.
        } 
      }
\end{figure}

A major challenge of modern physics is to
understand the fascinating phenomena in strongly-correlated transition
metal oxides (TMOs), which emerge in the neighborhood of the Mott
insulator state.  Some preeminent examples that have gathered the interest
for almost 30 years are high temperature superconductivity, 
colossal magnetoresistance, heavy fermion physics and,
of course, the Mott metal-insulator transition itself~\cite{Imada1998}.
Significant theoretical progress was made with the
introduction of Dynamical Mean Field Theory (DMFT) and its combination
with \emph{ab initio} Density Functional methods (LDA+DMFT), which allows
treatment of the interactions promoting itinerancy and localization of
electrons on equal footing~\cite{Georges1996,Kotliar2004,Kotliar2006}. 
Among the most emblematic achievements of DMFT is the prediction of a Hubbard
satellite, which splits off of the conduction band
of a metal.  This satellite results from the partial localization of
conduction electrons due to their mutual Coulomb repulsion.  Early
DMFT studies also showed that it is the precursor of
the localized electronic states of a Mott insulator~\cite{Zhang1993}.
Since then, these predictions promoted a large number of studies using
photoemission spectroscopy, which is a technique to directly
probe the presence of Hubbard bands. In this context, the TMO
system SrVO$_3$ has emerged as the \emph{drosophila} model system to
test the predictions of strongly correlated electron theories. In
fact, SrVO$_3$ is arguably the simplest correlated metal. It is a
simple cubic perovskite, with nominally one electron per V site, which
occupies a 3 fold degenerate $t_{2g}$ conduction band. While the
presence of a satellite in the photoemission spectra of Ni metal was
already well known, in the context of correlated TMOs, the Hubbard
band was originally reported in a systematic investigation of
Ca$_{1-x}$Sr$_x$VO$_3$~\cite{Inoue1995}, which was
followed by many subsequent studies, 
including ARPES~\cite{Takizawa2009,Yoshida2010,Aizaki2012} and
comparison with theoretical predictions 
(see for instance Refs.~\onlinecite{Rozenberg1996,Sekiyama2004, Pavarini2004,Amadon2008,
Aichhorn2009,Karolak2010,Lee2012,Taranto2013,Tomczak2014,vanRoekeghem2014} among others).

One of the most salient features in SrVO$_3$ is the
observation of a broad peak at an energy of about $-1.5$~eV in
angle integrated photoemission spectra,
(upper black curve in Fig.~\ref{fig:CompSTO-KTO-TiO2-SVO}), which
is interpreted as a Hubbard satellite linked to the V $t_{2g}$ electrons.  
This feature is also seen in a large range of $3d^1$ 
materials~\cite{Fujimori1992,Morikawa1996}.
The ratio of spectral strength between the quasiparticle state 
and the incoherent satellite in SrVO$_3$ 
is an important indicator of the magnitude 
of electron correlations~\cite{Georges1996,Imada1998}.
However, photoemission experiments using different photon energies 
or light brilliance have reported very dissimilar values 
for such ratio~\cite{Sekiyama2004}, 
making the quantitative benchmarking of realistic \emph{ab-initio}
theories for correlated-electron systems difficult
~\cite{Inoue1995,Sekiyama2004,Nekrasov2005,Eguchi2006,Taranto2013}.
Moreover, as shown in Fig.~\ref{fig:CompSTO-KTO-TiO2-SVO}, 
a broad peak at about the same energy is
also observed in several $d^0$ TMO cubic perovskites, such as SrTiO$_3$, KTaO$_3$, 
or anatase TiO$_2$. Nevertheless, in all these cases the
feature has been clearly linked to the presence of oxygen
defects~\cite{Aiura2002,Santander-Syro2011,Meevasana2011,King2012,Moser2013,
  Walker2014,Roedel2015,Walker2015}.
Interestingly, recent \emph{ab initio} calculations show that spectral
weight at $-1.3$~eV in SrTiO$_3$ most likely is not of Ti $t_{2g}$
orbital character, but should be understood as an in-gap defect
state with Ti $e_g$ character~\cite{Lin2012,Jeschke2015,Altmeyer2015,Lechermann2014}.
Thus, we are confronted with the fact that at about $1.5$~eV below the
Fermi level, we find the lower Hubbard bands of $d^1$ systems as well
as the in-gap states of oxygen-deficient $d^0$ systems.
In view of these observations one may unavoidably wonder (and worry), despite the great
success of DMFT methods, whether the putative Hubbard satellite of
SrVO$_3$  might also originate from oxygen vacancies states. 
Moreover, one should also worry about the possibility of these extrinsic states
affecting the features of the conduction band dispersion.

In the present work we resolve these issues in a thorough manner.
We  present a systematic photoemission study of SrVO$_3$, to
demonstrate dramatic consequences in the spectra due to production of oxygen vacancies.  
Using ARPES, we directly show that the UV/X rays used for measurements can
produce a large enhancement, of almost an order of magnitude,
of the peak at $-1.5$~eV, similar to the
effect observed in $d^0$ oxide insulators ~\cite{Santander-Syro2011,
Meevasana2011,King2012,Santander-Syro2012,Roedel2015}. 
Despite these significant effects on the energy states
around the Mott-Hubbard band, we are able to determine the {\it intrinsic} 
bulk SrVO$_3$ photoemission spectrum. 
We find that when the presence of oxygen vacancies is avoided, 
a clear signal of the correlated Hubbard band remains.
We support the interpretation of the experimental data by
means of state-of-the-art LDA+DMFT calculations on SrVO$_3$ \emph{with oxygen vacancies}.
Consistent with our experimental data, the calculations show that
oxygen vacancies produce states (of $e_g$ symmetry) at energies near the Hubbard satellite. 
While our study provides definite evidence of a correlated Hubbard band in SrVO$_3$ 
as predicted by DMFT, it also underlines the significant effects due to
oxygen vacancies, which may also affect photoemission data in other TMOs.

Figure~\ref{fig:SVO-vs-STO-UVdose}(a) shows the integrated
photoemission spectra of SrVO$_3$ as a function of the UV dose,
measured at the CASSIOPEE beamline of SOLEIL Synchrotron
under the same conditions of UV light brilliance
(about $\approx 5 \times 10^{9} \text{ photons s}^{-1} \mu\text{m}^{-2}$)
of any standard ARPES experiment at a third-generation synchrotron.
The measurements were done by continuously irradiating the sample with
$h\nu = 33$~eV photons while recording the spectra as a function of
irradiation time, with an accumulation time of about 2~minutes per
spectrum.  The blue and black curves show spectra for the lowest and
highest measured doses, obtained respectively after $\sim 2$~minutes
and $\sim 2$~hours of irradiation.
These data clearly demonstrate that the very UV/X rays used for
photoemission experiments can effect radical changes in the measured
spectra of SrVO$_3$. Note in fact that a similar effect
has been observed for VO$_2$~\cite{Muraoka2014}.
In particular, from Fig.~\ref{fig:SVO-vs-STO-UVdose}(a) we observe
that the amplitude of the in-gap state at $-1.5$~eV, and more
significantly, the ratio of in-gap to quasiparticle (QP) amplitudes,
strongly increase with increasing UV dose,
going from about $1:3$ in a pristine sample to more than $2:1$ 
in a heavily irradiated sample.
Importantly, note that the QP peak position remains basically
dose-independent, implying that the carriers created by the UV/X
irradiation do not significantly dope the conduction band, and form
dominantly localized states.
This is confirmed in Fig.~\ref{fig:SVO-vs-STO-UVdose}(b), which shows
that the Fermi momenta of the QP band, given by the peaks' positions
in the momentum distribution curves (MDCs) at $E_{\text{F}}$, are also
dose-independent.
Additional data presented in the Supplemental Material
further demonstrate that our measurements yield the
expected 3D bulk Fermi surface of SrVO$_3$.
Thus, the observed increase in intensity of the in-gap state upon UV
irradiation \emph{cannot} be ascribed to a change in filling of the
conduction band, which could have affected the electron correlations.
Instead, this unambiguously shows the light-assisted formation of
localized defect states at essentially the same energy as
that of the expected intrinsic lower Hubbard band --which should then
resemble the in-gap peak observed at the lowest UV doses.

In fact, as mentioned previously, it is well established that strong
doses of UV/X rays create a large concentration of oxygen vacancies in
several $d^0$ perovskites~\cite{Aiura2002,Santander-Syro2011,Meevasana2011,
King2012,Moser2013,Walker2014,Roedel2014,Walker2015,Roedel2015}.
As illustrated in Figs.~\ref{fig:SVO-vs-STO-UVdose}(c,~d) for the case
of SrTiO$_3$, the progressive doping of the surface region with oxygen vacancies,
due to synchrotron UV irradiation, has two effects: 
formation of a very intense in-gap state at about $-1.3$~eV, and, in contrast to SrVO$_3$, 
simultaneous creation of a sharp QP peak at $E_{\text{F}}$ corresponding
to a confined quasi-2D electron gas (2DEG) at the samples' surface.
The effective mass of such 2DEG, precisely determined by ARPES, 
matches the mass expected from density functional theory 
calculations~\cite{Santander-Syro2011,Meevasana2011,Shen2012,Plumb2014}. 
Thus, as in SrVO$_3$, the increase in intensity of the in-gap state observed
in SrTiO$_3$ upon UV/X irradiation cannot be due to an onset or increase of
electron correlations, and should be ascribed to an extrinsic effect.

We therefore conclude that, in SrVO$_3$, exposure to synchrotron UV/X
rays creates oxygen vacancies, which are in turn responsible for the
extrinsic increase in intensity of the in-gap state evidenced by our
measurements.
This effect, never discussed or taken into account before, can
seriously obscure the determination of the intrinsic spectral function
of this model system, thus hampering the advancement of valid theories
for correlated-electron systems.

All the previous findings imply that the correct experimental
determination of the \emph{intrinsic} spectral function of SrVO$_3$
should \emph{(i)} use samples that from the beginning have the lowest possible
concentration of oxygen vacancies, and \emph{(ii)} use doses of UV or X-ray light 
low enough to avoid any change in the measured spectra.
To this end, we measured bulk crystalline thin-films of
SrVO$_3$/Nb:SrTiO$_3$ grown \emph{in-situ}, in a pulsed laser deposition
(PLD) chamber directly connected to the ARPES setup,
at the beamline 2A of KEK-Photon Factory (KEK-PF)~\cite{Yoshimatsu2010,Aizaki2012}.
The PLD growth was performed under a pressure below $10^{-7}$~Torr, 
to obtain an UHV-clean surface, using a Sr$_2$V$_2$O$_7$ target, 
which has excess oxygen with respect to SrVO$_3$,
thus minimizing the formation of vacancies during the growth
--see the Supplemental Material for further details. 
Furthermore, the UV light brilliance used in these experiments was
about $100$ times lower than the one in Figs.~\ref{fig:SVO-vs-STO-UVdose}(a,~b)
from measurements at SOLEIL.
We checked (Supplemental Material) that under these conditions
the spectra did not change with time, even after several hours of
measurements.
The resulting energy-momentum ARPES map is presented in
Fig.~\ref{fig:SVO-vs-STO-UVdose}(e).  
One clearly observes the dispersion in the QP band 
along with the weaker intensity in-gap state, the intrinsic lower Hubbard band,
similar as previously reported~\cite{Takizawa2009}.
By contrast, Fig.~\ref{fig:SVO-vs-STO-UVdose}(f) shows the
momentum-resolved electronic structure of a sample, 
measured at SOLEIL, that was intensively irradiated. 
There, the peak at $-1.5$~eV becomes broader,
more intense, and non-dispersive --all characteristic signatures of a
high random concentration of oxygen vacancies. 

\begin{figure}[tb]
        \includegraphics[clip, angle=270, width=0.45\textwidth]{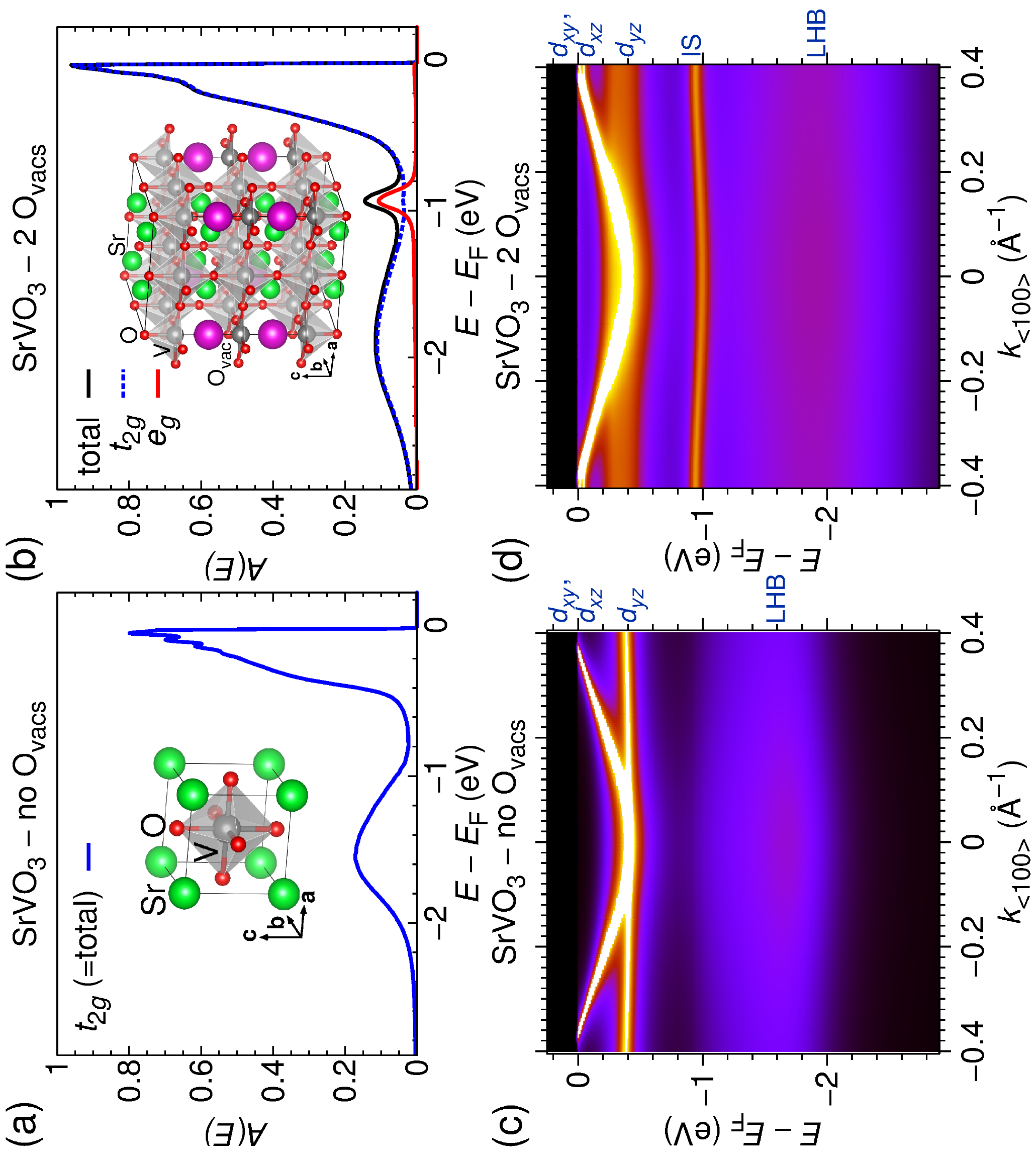}
	\caption{(Color online) 
    LDA+DMFT results for {\srvo}
    including bandwidth renormalization effects~\cite{Casula2012}. 
	(a) $k$-integrated spectral function for bulk {\srvo}.  
	The V $t_{2g}$ orbitals show a quasiparticle peak at $E_{\text{F}}$ 
	and a lower Hubbard band at $-1.6$~eV. 
	(b) Spectral function for the $2 \times 2 \times3$ supercell of {\srvo} 
	with two oxygen vacancies. An additional
	non-dispersive V $e_g$ vacancy state originating from the V atom
	neighboring the oxygen vacancies lead to a sharp peak below the
	Fermi level at $-1.1$~eV.  The V $t_{2g}$ orbitals show a 
	quasiparticle peak at $E_{\text{F}}$ and a lower Hubbard band at $-1.8$~eV.
	(c) and (d) show the corresponding spectral functions 
	(multiplied by a Fermi-Dirac function at $20$~K)
	along the X-$\Gamma$-X path. 
	}
\label{fig:2vac_dft_dmft}
\end{figure}

To rationalize from a microscopic point of view the influence of
oxygen vacancies on the electronic structure of SrVO$_3$, 
we performed charge self-consistent LDA+DMFT calculations 
for bulk SrVO$_3$ and various relaxed
oxygen-deficient SrVO$_3$ supercells. 
The latter are computationally demanding calculations. We
shall focus here on the case of a $2\times 2\times 3$ supercell with
two oxygen vacancies located at opposite apical sites of one vanadium
atom, as shown in the inset of Fig.~\ref{fig:2vac_dft_dmft}~(b).
We use such vacancy arrangement as it is the prototypical one
for $d^0$ compounds~\cite{Shen2012}.

For our LDA+DMFT calculations we chose values of
$U=2.5$~eV and $J=0.6$~eV for vanadium and included the effects 
of bandwidth renormalization due to dynamically screened Coulomb interactions 
by following the prescription suggested in Ref.~\cite{Casula2012}
(the LDA+DMFT unrenormalized data are shown in the Supplemental Material).
In Figs.~\ref{fig:2vac_dft_dmft}(a) and (c) we show, respectively, the
results  of the $k$-integrated and $k$-resolved spectral functions
for bulk SrVO$_3$ without oxygen vacancies.  We find the expected
features of a $t_{2g}$ quasiparticle peak at the Fermi level and a
lower Hubbard band at negative energies of the same $t_{2g}$ nature,
in agreement with the photoemission spectra 
in Figs.~\ref{fig:SVO-vs-STO-UVdose}(a) and (e). 
The light band at $E_{\text{F}}$ along $k_{<100>}$, Fig.~\ref{fig:2vac_dft_dmft}(c), 
consists of two degenerate bands of $d_{xy}$ and $d_{xz}$ characters, 
while the heavy band along the same direction has $d_{yz}$ character.
While comparing  with the measured $k$-resolved spectral function, 
Fig.~\ref{fig:SVO-vs-STO-UVdose}(e), one should bear in mind 
that along $\Gamma$-X (or $\Gamma$-Y) the heavy $d_{yz}$ (or $d_{xz}$) bands
are silenced by dipole-transition selection rules in the experiment~\cite{Santander-Syro2011}. 
Inclusion of bandwidth renormalization~\cite{Casula2012} renders
the lower Hubbard band at an energy ($-1.6$~eV) 
in reasonable agreement with experiment ($-1.5$~eV).
We adopted values for $U$ and $J$ from the literature, and did not try
to optimize the quantitative agreement with the experimental data.
As we show below, in the calculations with oxygen vacancies 
the value used for $U$ facilitates the visualization
of the contributions from the Hubbard and localized states to
the  incoherent peak at $\sim -1.5$~eV.

The removal of oxygen atoms in the system leads to the donation of two
electrons per oxygen to its surrounding.  Already at the level of
density functional theory (DFT) in the local density approximation
(LDA) (see Supplemental Material), 
we find that most of the charge coming from
the additional electrons is transferred to the $3d_{z^2}$ orbitals of
the neighboring V atom, developing into a sharp peak of $e_g$ symmetry
located around $-1.1$~eV, {\it i.e.} at an energy close to the
position of the experimentally observed oxygen vacancy states.
In analogy to the experimental average over many lattice sites,
note that averaging among various supercells with different 
oxygen vacancy locations and concentrations (what is beyond the scope of the present work) 
would result in a wider in-gap $e_g$ band, 
as demonstrated for the case of SrTiO$_3$ 
--see Fig. 3 of Ref.~\onlinecite{Jeschke2015}.
By including electronic correlations within (bandwidth renormalized) LDA+DMFT 
we then see that all the experimental observations qualitatively emerge. 
In fact, the conducting $t_{2g}$ orbitals develop a lower Hubbard band 
peaked at energies about $-1.8$~eV (Fig.~\ref{fig:2vac_dft_dmft}(b) and (d)) 
similar to the bulk case without oxygen vacancies. 
Most notably, this lower Hubbard satellite 
does not increase in amplitude with the introduction of vacancies, 
but rather broadens. In addition, the oxygen-vacancy defect
states situated at about $-1$~eV 
remain qualitatively unchanged by the correlation effects, but
experience a broadening with respect to the pure LDA case.
This is in agreement with the photoemission data, evidencing that the increase in intensity 
of the in-gap state in the oxygen-deficient SrVO$_3$ is not to be attributed 
to an increase in population of the lower Hubbard satellite, 
but instead to the manifestation of vacancy states of $e_g$ character.

In summary, we performed a detailed study of the effects of oxygen vacancies
in the spectroscopy of the archetypal strongly correlated electron system SrVO$_3$.  
We found that oxygen vacancy states, which are created by UV/X-ray irradiation,
occur at energies close to the Hubbard satellite.
This dramatically affects the measured line-shape of the Mott-Hubbard band
and the ratio of intensities between the quasi-particle and the Mott-Hubbard peaks.
By means of a systematic study under controlled irradiation dose
and avoiding the formation of oxygen vacancies,
we were able to obtain the intrinsic
occupied spectral function of the bulk SrVO$_3$ system. 
Our experimental interpretation is supported by
LDA+DMFT calculations, which provided further insight on the likely
nature of the oxygen vacancy states.

\acknowledgments 
We thank Silke Biermann, Ralph Claessen, Marc Gabay and Michael Sing for discussions.
This work was supported by public grants from the French National Research Agency (ANR), 
project LACUNES No ANR-13-BS04-0006-01, 
and the ``Laboratoire d'Excellence Physique Atomes Lumi\`ere Mati\`ere'' 
(LabEx PALM project ELECTROX) overseen by the ANR as part of the 
``Investissements d'Avenir'' program (reference: ANR-10-LABX-0039).
S.B., A.J.K., F.L., H.O.J. and R.V. gratefully acknowledge the Deutsche
Forschungsgemeinschaft for financial support through grant FOR 1346.
T.~C.~R. acknowledges funding from the RTRA--Triangle de la Physique (project PEGASOS).
A.F.S.-S. thanks support from the Institut Universitaire de France.

\section{Supplemental Material}

\subsection{Experimental Methods}
\subsubsection{Sample preparation}
The SrVO$_3$ thin films measured at SOLEIL were grown at the GEMaC
laboratory, onto atomically flat TiO$_2$-terminated (100) SrTiO$_3$ substrates 
by pulsed laser deposition (PLD) under a partial oxygen pressure of 
$5 \times 10^{-6}$~Torr 
at temperature of $750^{\circ}$C. A 248~nm wavelength KrF excimer laser was employed 
with a repetition rate of 1~Hz and a fluency of 1.9~J/cm$^2$. 
During the growth, surface structure was characterized by 
Reflection High-Energy Electron Diffraction (RHEED). 
After the growth, oxygen pressure was reduced to 
$5 \times 10^{-8}$~Torr 
when the sample was cooled to room temperature. 
Surface morphology was carried out with atomic force microscopy (AFM, Bruker Dimension 3100) 
in tapping mode and the Root Mean Square (rms) roughness of the film was 0.4~nm. 
To clean the surfaces in UHV prior to ARPES experiments at SOLEIL, 
the SrVO$_3$ thin films were annealed at a temperature 
$T=550$~$^\circ$C for $t=5-20$~min at pressures lower than 
$p < 2 \times 10^{-8}$~Torr. 
One of the samples was Ar$^+$ sputtered ($U = 1000$~V, $t = 7$~min) prior to the UHV annealing,
without noticeable changes in the ARPES data.
After the UHV annealing, the quality of the surfaces was confirmed 
by low-energy electron diffraction (LEED). 

The SrVO$_3$ films measured at KEK were grown \emph{in-situ}, 
on a PLD chamber directly accessible from the ARPES UHV chamber, 
on single-crystalline 0.05~wt\% Nb-doped SrTiO$_3$~$(001)$ substrates. 
The substrates were annealed at $1050^{\circ}$C under an oxygen pressure of 
$\sim 8\times 10^{-8}$~Torr 
to obtain atomically flat TiO$_2$-terminated surfaces. 
The SrVO$_3$ thin film was deposited on the substrate at $850^{\circ}$C,
under a high vacuum below $\sim 10^{-7}$~Torr, to obtain an UHV clean surface,
using a Sr$_2$V$_2$O$_7$ target, which has excess oxygen with respect to SrVO$_3$,
and thus minimizes the formation of vacancies during the growth.
LEED measurements confirmed clean cristalline SrVO$_3$ films
with a $c(4 \times 4)$ surface reconstruction.
The surface morphology of the measured thin films was confirmed by \emph{ex-situ} 
atomic force microscopy, showing atomically flat step-and-terrace structure. 
The coherent growth of these SrVO$_3$ thin film was confirmed by four-circle 
X-ray diffraction measurements. The characterization results are essentially identical 
to those reported in previous studies~\cite{Yoshimatsu2010,Yoshimatsu2011,Kobayashi2015},
and indicate that there are no detectable structural disorders in the prepared SrVO$_3$ thin films 
grown on Nb:SrTiO$_{3}$~$(001)$ substrates.

\subsubsection{ARPES measurements}
\begin{figure*}[t]
    \begin{center}
        \includegraphics[clip, width=14cm]{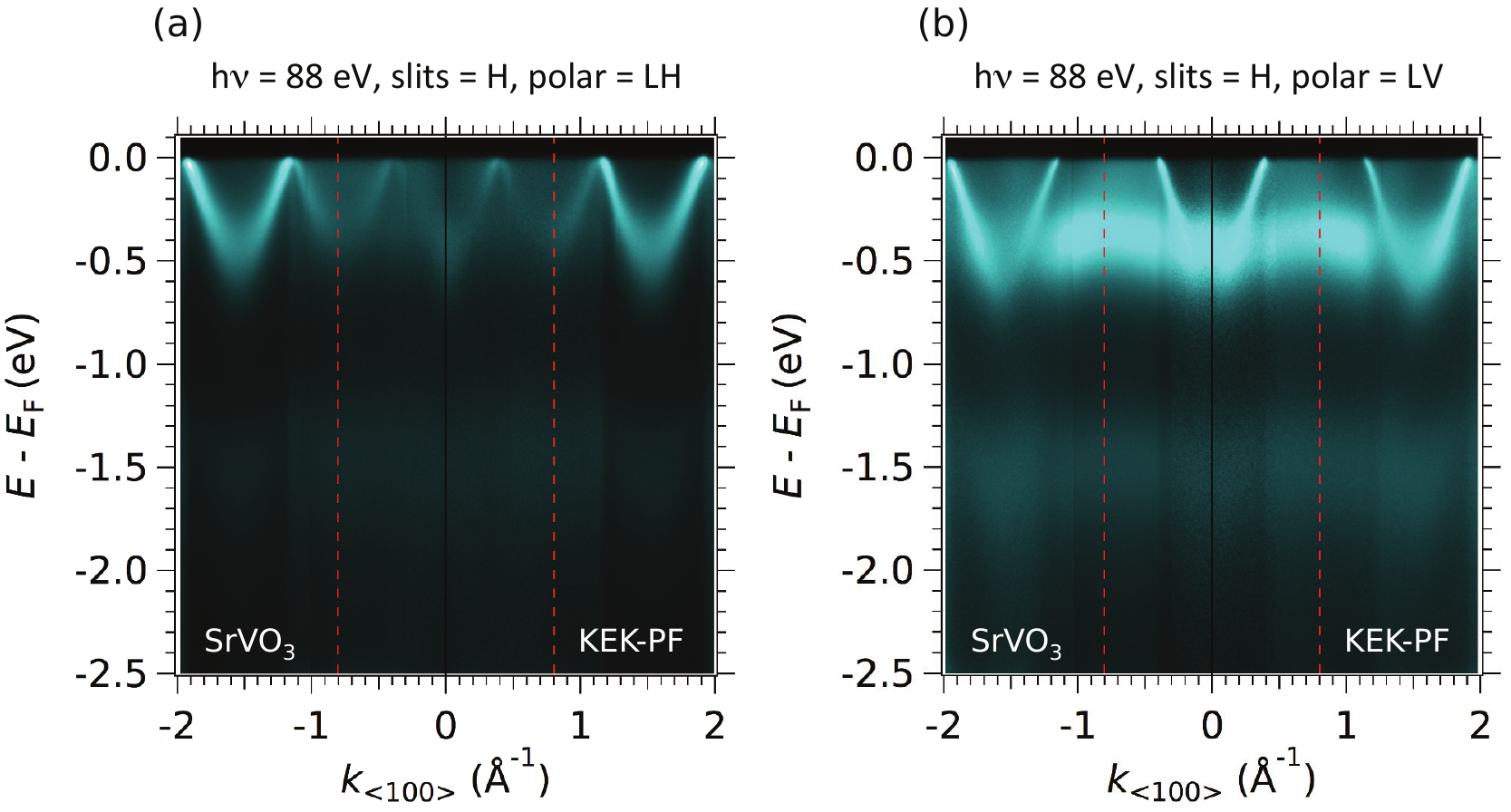}
    \end{center}
    \caption{\footnotesize{
        	Energy-momentum ARPES intensity maps spanning 3 consecutive Brillouin zones
        	on an SrVO$_3$ thin film prepared \emph{in-situ}. 
        	Vertical dashed red lines show the edges of the 
        	bulk unreconstructed Brillouin-zone.
            The data were measured at KEK-PF using $h\nu = 88$~eV photons 
        	with (a) linear horizontal and (b) linear vertical polarizations, 
        	and a hemispherical electron analyzer with horizontal slits. 
        	The sample temperature was 20~K.
        	Surface umklapp bands induced by the surface reconstruction are best observed
        	in the data using linear horizontal polarization.
        	The spectra taken with linear vertical photons reveal the light and heavy bands
        	of the intrinsic bulk electronic structure of SrVO$_3$. 	
        	}
        	}
    \label{fig:Surface-Umklapps}     
\end{figure*}

The ARPES measurements were conducted at the CASSIOPEE beamline 
of Synchrotron SOLEIL (France), and at beamline 2A of KEK-Photon Factory (KEK-PF, Japan).
We used linearly polarized photons in the energy range $30-110$~eV and 
hemispherical electron analyzers with vertical slits at SOLEIL 
and horizontal slits at KEK-PF.
The angular and energy resolutions were $0.25^{\circ}$ and 15~meV. 
The mean diameter of the incident photon beam was smaller than 100~$\upmu$m.
The UV light brilliance, 
measured using calibrated photodiodes,
was $\approx 5 \times 10^{9} \text{ photons s}^{-1} \mu\text{m}^{-2}$
at SOLEIL, and about 100 times smaller at KEK-PF.
The (001) oriented SrVO$_3$ samples were cooled down to $T=20$~K before measuring.
Unless specified otherwise, all data was taken at that temperature.
Measuring at such low temperature minimizes any possible diffusion of oxygen vacancies
into the bulk.
The results have been reproduced on more than 5 samples. 
All through this paper, directions and planes are defined 
in the cubic unit cell of SrVO$_3$.
We note $[hkl]$ the crystallographic directions in real space, 
$\langle hkl \rangle$ the corresponding  directions in reciprocal space, 
and $(hkl)$ the planes orthogonal to those directions. 
The indices $h$, $k$, and $l$ of $\Gamma_{hkl}$ correspond to
the reciprocal lattice vectors of the cubic unit cell of SrVO$_3$.

\subsection{Surface umklapp bands in thin-films of $\textrm{SrVO}_3$}
The surface reconstruction in our SrVO$_3$ thin films leads to the occurence 
of surface umklapp bands in the ARPES spectra, as shown in Figs.~\ref{fig:Surface-Umklapps}(a,~b)
for data taken with linear horizontal and linear vertical photons polarizations, respectively.
These extra bands, best observed with linear horizontal photons, 
result simply from the folding of the bulk bands at the surface due to the
superposition of the bulk and the reconstructed surface periodicities.
Their presence is a final-state effect and does not affect 
the bulk electronic structure or the effects due to oxygen vacancies
discussed in this work.
Note that the spectra taken with linear vertical photons show both the light and heavy bands
expected for the intrinsic bulk electronic structure of SrVO$_3$, discussed in the next section.

\subsection{Bulk electronic structure of $\textrm{SrVO}_3$}
\begin{figure}[h]
    \begin{center}
        \includegraphics[clip, width=8cm]{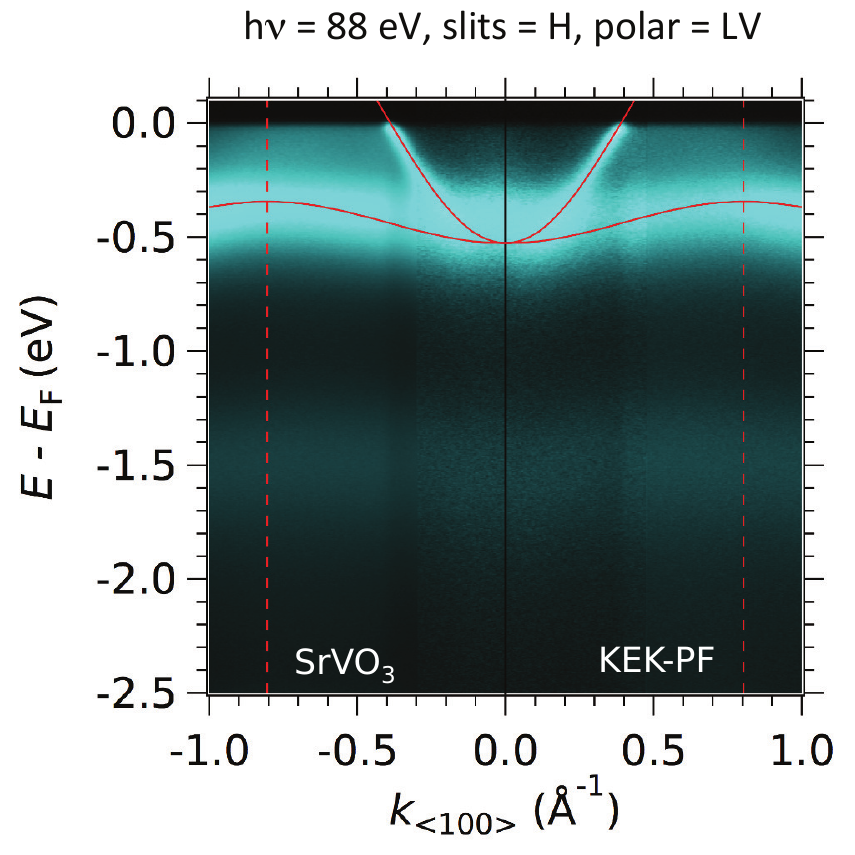}
    \end{center}
    \caption{\footnotesize{
        	Zoom around $\Gamma$ of the energy-momentum ARPES intensity map
        	shown in Fig.~\ref{fig:Surface-Umklapps}(b), measured
        	at KEK-PF using $h\nu = 88$~eV photons 
        	with linear vertical polarization 
        	and a hemispherical electron analyzer with horizontal slits. 	
        	Vertical dashed red lines show the Brillouin-zone edges.
        	Continuous red curves are cosine fits to the bands.
        	}
        	}
    \label{fig:SVO-Light-Heavy-Bands}     
\end{figure}

\begin{figure*}[t]
    \begin{center}
        \includegraphics[clip, width=16cm]{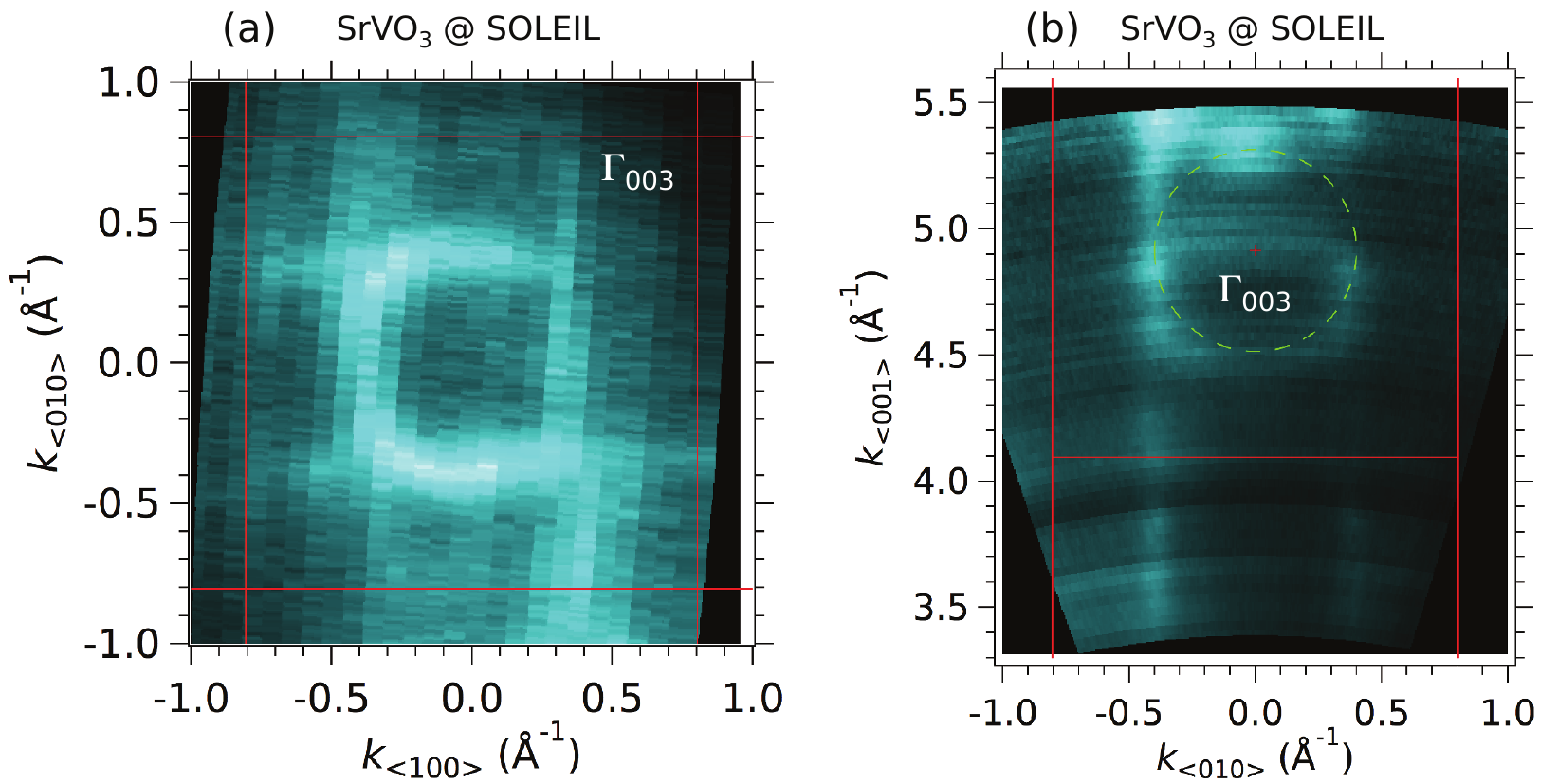}
    \end{center}
    \caption{\footnotesize{
        	(a)~In-plane Fermi-surface of SrVO$_3$(001), measured 
        		using linear horizontal photons at $h\nu = 72$~eV, 
        		and a hemispherical electron analyzer with vertical slits.
        	(b)~Out-of-plane Fermi surface of SrVO$_3$(001), measured 
        		using linear vertical photons 
        		from $30$~eV to $95$~eV, in steps of 1~eV, 
        		and a hemispherical electron analyzer with vertical slits.  
        		The green dashed circle is as guide to the eye showing
        		the quasi-circular Fermi-surface formed by the $d_{yz}$ states.
        		In both panels, the red lines delimit the Brillouin-zone.
        		All data from this figure were measured at CASSIOPEE (SOLEIL),
        		using the same light brilliance
        		conditions of Figs.~\ref{fig:SVO-vs-STO-UVdose}(a,~b).
        	}
        	}
    \label{fig:FS-SrVO3}     
\end{figure*}

Figure~\ref{fig:SVO-Light-Heavy-Bands} presents a zoom around $\Gamma$ of the 
ARPES energy-momentum map shown in Fig.~\ref{fig:Surface-Umklapps}(b),
which was taken using linear vertical photons and a hemispherical electron analyzer 
with horizontal slits. This configuration of experimental geometry and light polarization
allows observing both the light and heavy conduction bands of SrVO$_3$.
We fit these bands using simple cosine dispersions of the form
$E(k) = E_0 - 2t\cos(ka)$, where $E_0$ is a constant, $4t$ is the bandwidth,
and $a$ is the in-plane lattice parameter of the SrVO$_3$ thin films 
($a=3.905$~\AA~in the case of films grown on SrTiO$_3$).
For small $ka$, each cosine band can be approximated by a free-electron-like parabola
of effective mass $m^{\star}$ given by $ta^2 = \hbar^2/(2m^{\star})$.
The best fits, shown by the continuous red curves in Fig.~\ref{fig:SVO-Light-Heavy-Bands},
yield effective masses $m_l^{\star} = 0.9 m_e$ and $m_h^{\star} = 5.5 m_e$ 
for the light and heavy bands, respectively ($m_e$ is the free electron mass).

Figures~\ref{fig:FS-SrVO3}(a,~b) show the ARPES Fermi surface maps of SrVO$_3$
cut, respectively, along the $k_{x}-k_{y}$ and $k_{y}-k_{z}$ planes 
(or in-plane and out-of-plane).
As was shown in Figs.~\ref{fig:SVO-vs-STO-UVdose}(a,~b), the Fermi momenta and band filling 
of the QP conduction band are independent of the UV dose. 
This ensures that the Fermi surface maps are exempt of extrinsic spurious effects 
due to the UV-assisted creation of oxygen vacancies.

Our measured Fermi surface agrees with previous ARPES reports
on the electronic structure of SrVO$_3$~\cite{Takizawa2009,Yoshida2010,Aizaki2012}.
Furthermore, the bulk 3D character of our measured Fermi surface
is demonstrated by the observation, in Fig.~\ref{fig:FS-SrVO3}(b),
of dispersive $d_{yz}$ states forming a quasi-circular Fermi sheet 
in the $k_{y}-k_{z}$ plane (green dashed circle serving as guide to the eye).

The 3D density of carriers in the bulk conduction band ($n_{3D}$) can be directly calculated
from the volume enclosed by the measured Fermi surface ($V_F$) as $n_{3D} = V_F/4\pi^3$.
In very good approximation, $V_F$ corresponds to the volume of three mutually orthogonal
inter-penetrating cylinders of cross-sectional Fermi radius $k_F = 0.4$~\AA$^{-1}$ 
and length $2\pi/a$ ($a$ is the cubic lattice constant of SrVO$_3$), \emph{i.e.}:
$V_F \approx 3\times \pi k_F^2 \times 2\pi/a - 2 \times (4/3)\pi k_F^3$,
where the last term avoids counting three times the inner Fermi quasi-sphere 
at the intersection of the three cylinders. 
This gives $n_{3D} \approx 0.9 e^-/a^3$ ($a^3$ is the unit-cell volume), 
which is in excellent agreement with the expected value of 1 electron per Vanadium.
This again confirms that our ARPES measurements probe the 3D bulk Fermi surface.

\subsection{Electronic structure of vacancy-free $\textrm{SrVO}_3$ using ultra-low dose UV light}
\begin{figure}
    \begin{center}
        \includegraphics[clip, width=6cm]{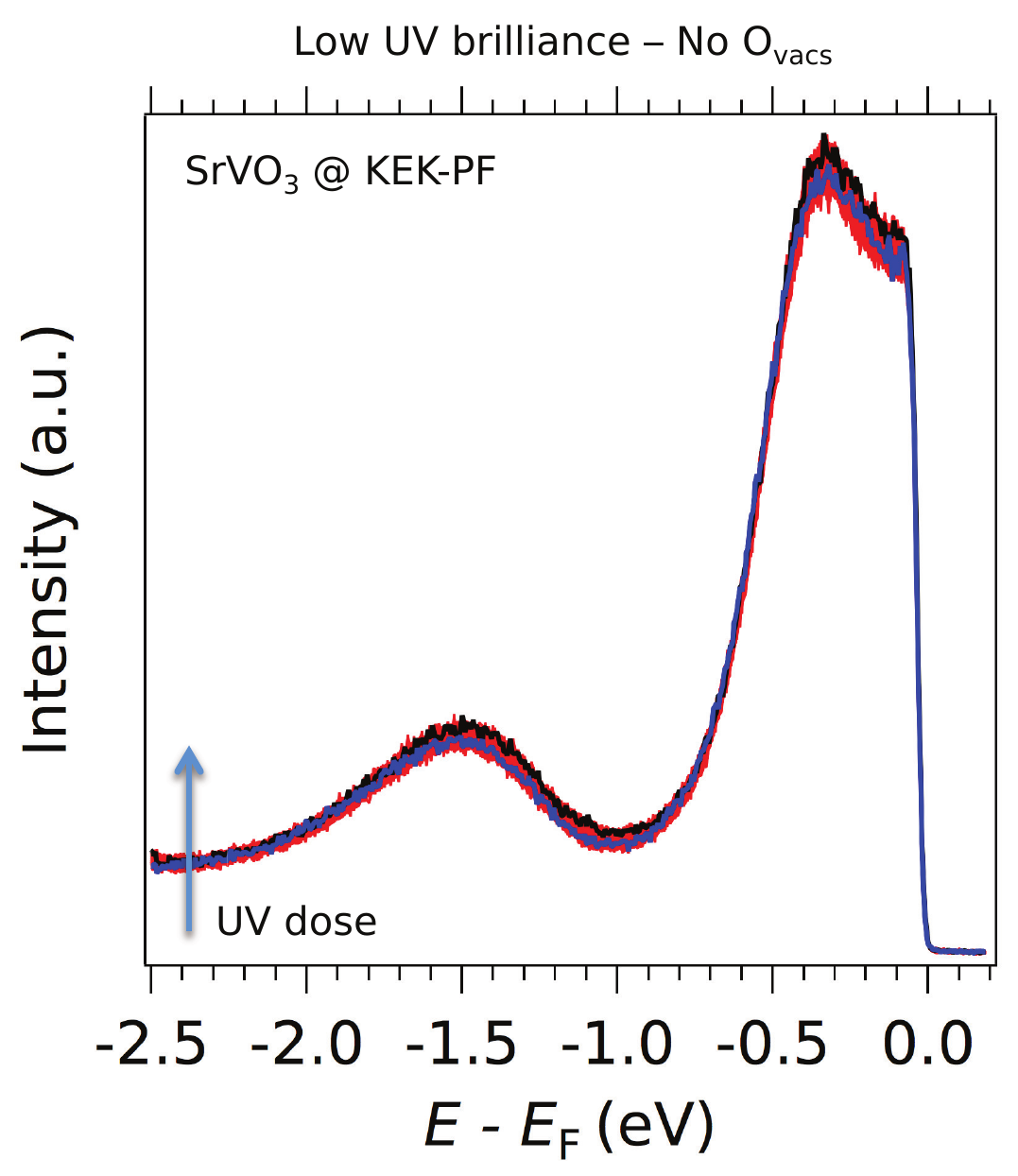}
    \end{center}
    \caption{\footnotesize{
        	Photoemission spectra of SrVO$_3$ as a function of irradiation time
        	using an ultra-low dose of UV light. 
        	The sample was prepared \emph{in-situ}.  
        	This eliminates the need to re-anneal in vacuum to clean the surface
        	prior to measurements, 
        	and thus minimizes the formation of oxygen vacancies.
        	The energy distribution curves (EDCs) were extracted from ARPES data 
        	around the $\Gamma_{003}$ point 
        	integrated along the $k = <100>$ direction.
        	The blue and black curves show spectra for the lowest and highest measured doses, 
        	obtained respectively after $\sim 2$~minutes and $\sim 2$~hours of irradiation.
        	The data from this figure were measured at KEK-PF using $h\nu = 88$~eV photons 
        	with linear horizontal polarization, and a hemispherical electron analyzer 
        	with horizontal slits.
        	}
        	}
    \label{fig:EDCs-SrVO3-KEK}     
\end{figure}

\begin{figure*}[t]
    \begin{center}
        \includegraphics[clip, width=14cm]{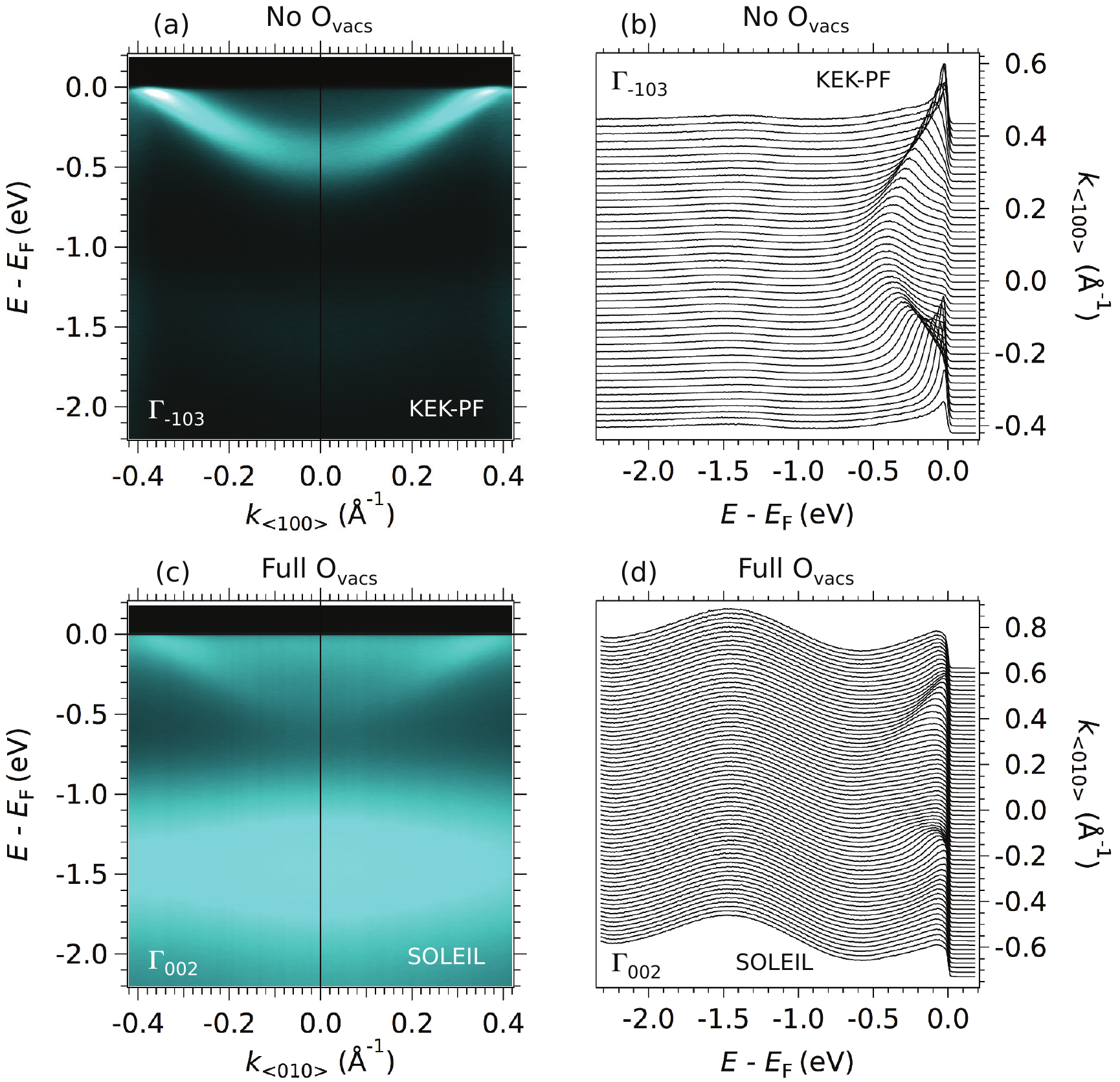}
    \end{center}
    \caption{\footnotesize{
        	(a)~Raw energy-momentum ARPES intensity map
        		measured at a low UV dose on an SrVO$_3$ sample prepared \emph{in-situ}.
        		The data were measured at KEK-PF using $h\nu = 88$~eV photons 
        		with linear horizontal polarization, and a hemispherical electron analyzer 
        		with horizontal slits.
        	(b)~Corresponding raw EDCs.
			(c,~d)~Same as (a,~b) after a strong UV irradiation. 
				This data were measured at CASSIOPEE (SOLEIL) using $h\nu = 33$~eV photons 
        		with linear vertical polarization, and a hemispherical electron analyzer 
        		with vertical slits.
               	All data were taken at 20~K.
        	}
        	}
    \label{fig:Raw-ARPES-UV-dose-SVO}     
\end{figure*}

Fig.~\ref{fig:EDCs-SrVO3-KEK} shows photoemission integrated spectra of a bulk 
SrVO$_3$/SrTiO$_3$ thin-film prepared and measured \emph{in-situ} at KEK-PF.
This eliminates the need to re-anneal in vacuum to clean the surface, 
and thus minimizes the formation of oxygen vacancies before the measurements.
Additionally, the spectra were measured using ultra-low doses of UV light.
As can be seen from the figure, this careful measurement protocol
prevents the formation of oxygen vacancies, allowing to acquire high-quality data
that shows no significant evolution with accumulation time,
even after several hours of measurements.
Thus, we define the obtained spectra as the ``intrinsic'' 
(occupied part of the) spectral function of SrVO$_3$.
Hence, the intrinsic ratio between the intensities of the lower Hubbard band and the QP peak
is $1:3$.

\subsection{Raw ARPES data: vacancy-free \emph{vs} vacancy-full $\textrm{SrVO}_3$}
Figs.~\ref{fig:Raw-ARPES-UV-dose-SVO}(a,~b) show the raw ARPES data for the QP and 
lower Hubbard band dispersions in a vacancy-free sample measured under low UV dose, 
corresponding to the second derivative of Fig.~\ref{fig:SVO-vs-STO-UVdose}(e).
Likewise, Figs.~\ref{fig:Raw-ARPES-UV-dose-SVO}(c,~d) 
show the raw ARPES data for the QP and in-gap state for a sample exposed
to a strong UV dose, corresponding to the second derivative of 
Fig.~\ref{fig:SVO-vs-STO-UVdose}(f).

As mentioned in the main text, in the vacancy-free sample 
the QP band and intrinsic lower Hubbard band show the same dispersion,
while in the sample that has been strongly irradiated, and has thus a large concentration of
oxygen vacancies, the peak at $-1.5$~eV becomes broader, 
more intense, and non-dispersive.

\subsection{Theoretical Methods}

\subsubsection{LDA and LDA+DMFT Calculations}
For the  DFT in the local density approximation (LDA) 
and LDA+DMFT calculations we consider
a $2\times 2 \times 3$ supercell where two adjacent oxygen atoms to a vanadium
have been removed. This leads to a stripe-like configuration of vacancies with a concentration
of $2/36\approx 5.56\%$. The internal atomic positions of this  structure has 
been relaxed using the GPAW code~\cite{gpaw}.

For the LDA+DMFT calculations (see Ref.~\cite{backes2014} for
a detailed explanation) we used the \textsc{WIEN2k}~\cite{Blaha01}
implementation in the local density approximation in combination
with a continuous-time quantum Monte Carlo (CTQMC)
impurity solver in the hybridization
expansion~\cite{Werner06} from the ALPS~\cite{ALPS11,Gull11a} project.
We projected the Bloch wave functions onto localized V $3d$
orbitals~\cite{Aichhorn2009,Ferber2014} in the unit cell, leading to a set
of 6 inequivalent impurities for the two vacancy structure. 
Within the DMFT approximation we assume that (i) the multiple impurity atoms
only hybridize with an effective bath that is determined self-consistently and (ii) 
the intersite hybridization to be small. This allows us to solve the impurity problems
separately and to write the full self-energy in Bloch space as
\begin{align}
	\Sigma_{\nu\nu'}(k,i\omega_n)
	&= \left( \sum_a  P^{\dagger}_a(k) \Sigma_a(i\omega_n) P_a(k) \right)_{\nu\nu'},
\end{align}
where the sum is performed over all impurity atoms $a$, with the corresponding
projection operators $P_a(k)$.
The projection comprises the V $t_{2g}$ and $e_g$ orbitals in the energy window
$[-1.5, 6.0]$~eV. 
The calculations were performed for the inverse temperature 
$\beta = 40\,\mathrm{eV^{-1}}$ ($290$ K) 
with the interaction parameters used in the
definition of the Slater integrals~\cite{Liechtenstein1995} 
with average $U=2.5\,\mathrm{eV}$ and $J_H=0.6\,\mathrm{eV}$.
For the double counting correction we used the FLL~\cite{Anisimov93, Dudarev98} scheme 
and we checked that different values of the double counting only lead to small quantitative 
changes in the spectral function.
The continuation of the Monte Carlo data to the real axis was done by stochastic
analytic continuation~\cite{beach2004}.

\subsubsection{Standard LDA+DMFT vs. LDA+DMFT including dynamical screening}
\begin{figure}[h]
	\begin{center}
		\includegraphics[clip,width=0.23\textwidth]{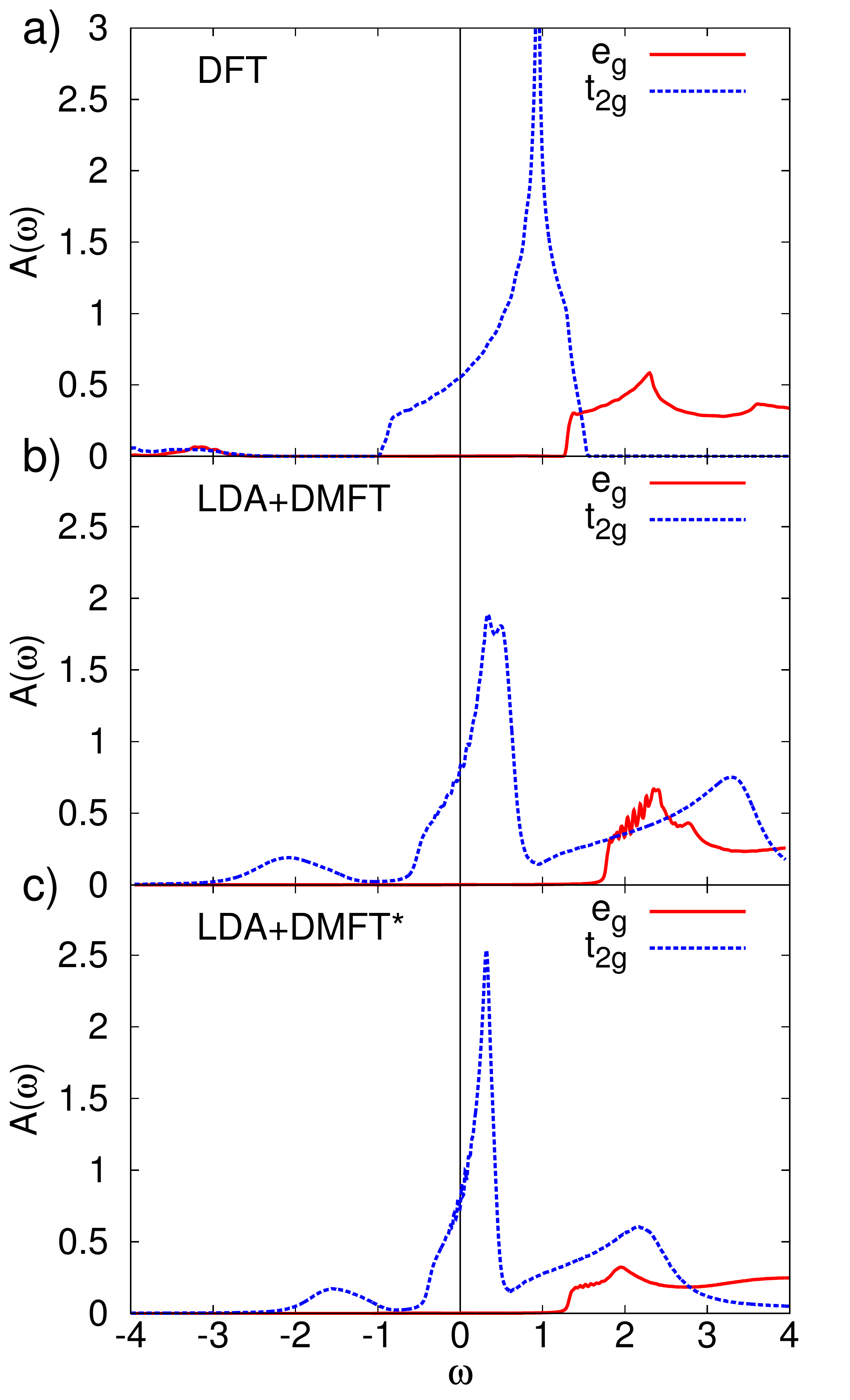}
        \includegraphics[clip,width=0.23\textwidth]{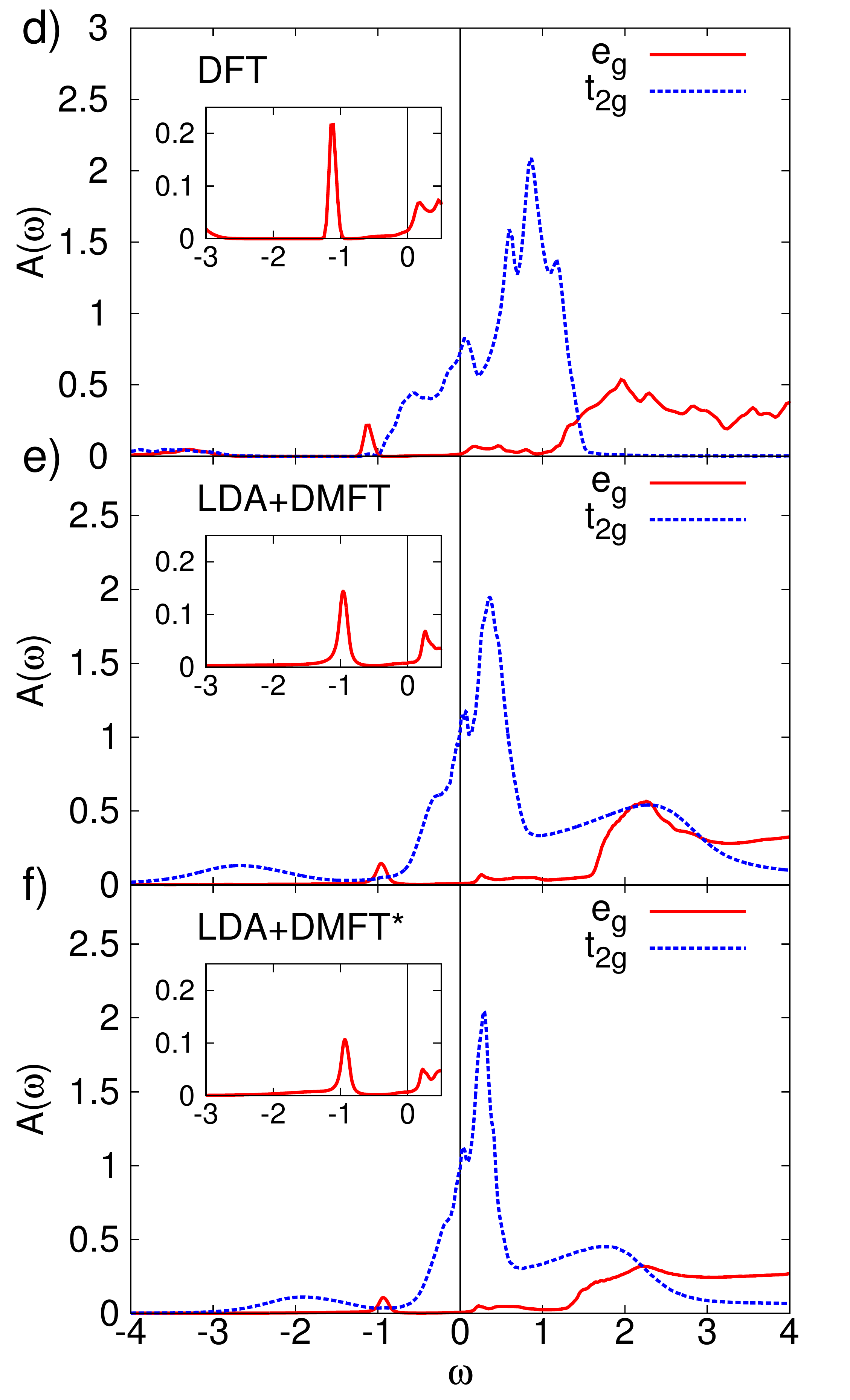}
	\end{center}
	\caption{(Color online) 
	Comparison for bulk SrVO$_3$ between (a) the DFT result, (b) the LDA+DMFT result,
        and (c) the LDA+DMFT including an effective treatment 
        of the dynamical Coulomb interacting screeening
        (explained in the main text) indicated by a *. 
	(d-f): Same comparison for the 2 vacancy structure of SrVO$_3$. The main effect of the
        renormalization factor is a shift of the upper and lower $t_{2g}$ Hubbard bands towards the
        Fermi level, yielding a better agreement with experiment. 
	}
\label{fig:dft_dmft_rescale_comp}
\end{figure}

In this work we included the effects of bandwidth renormalization 
due to dynamically screened Coulomb interactions 
using a low-energy effective model with an effective Hamiltonian
\begin{align}
H_{eff} 
&= -\sum\limits_{ij\sigma} Z_B t_{ij} d^{\dagger}_{i\sigma}d_{i\sigma}
+ U_0 \sum\limits_i d^{\dagger}_{i\uparrow}d_{i\uparrow} d^{\dagger}_{i\downarrow}d_{i\downarrow},
\end{align}
with the screened Hubbard interaction $U_0$. 
This approach has been suggested in Ref.~\cite{Casula2012}, and proved
to be a good approximation to the full treatment of the dynamically screened Coulomb interaction.
For the bandwidth renormalization factor we used $Z_B = 0.7$~\cite{Casula2012}.

In Fig.~\ref{fig:dft_dmft_rescale_comp} we show a comparison between standard DFT, 
``standard'' LDA+DMFT and the LDA+DMFT approach including the effective screening 
of the Coulomb interaction for a)-c) bulk SrVO$_3$ 
and d)-f) the two vacancy structure of SrVO$_3$. 
The main effect of including the Coulomb interaction screening
via this approach is a shift of the upper and lower $t_{2g}$ Hubbard bands 
towards the Fermi level, originating from the effective reduction of bandwidth.
Compared to experiment (as discussed in the main text), ``standard'' LDA+DMFT 
consistently locates the lower Hubbard band at higher binding energies,
in both the bulk and vacancy structure, whereas the effective
model yields a much better agreement. 
Especially in the bulk system the position of the lower Hubbard band
is brought to a good agreement. 

Finally, the calculations with oxygen vacancies 
produce a ladder of heavy bands near $E_{\text{F}}$, 
that originate from the non-equivalency in all the vanadium atoms 
of the super-cell contributing $t_{2g}$ bands near the Fermi level.
This V non-equivalency is due to the presence of oxygen vacancies  
that lower the symmetry.  
In the more realistic case with larger cells, only the in-gap $e_g$ states
of the V near the vacancy/vacancies would produce a significant spectral feature, 
while essentially all the V atoms of the cell would be equivalent, 
and the effects of the $t_{2g}$ states coming from the few 
non-equivalent vanadium atoms around the vacancy would be negligible. 
At present, such a calculation is however computationally unfeasible 
in the framework of LDA+DMFT.


\end{document}